\documentclass{osa-article1}

\journal{osajournal}

\articletype{Research Article}

\begin{document}

\title{Rayleigh fading suppression in one-dimension optical scatters}

\author{Shengtao Lin,\authormark{1} Zinan Wang,\authormark{1,2,*} Ji Xiong,\authormark{1} Yun Fu,\authormark{1} Jialin Jiang,\authormark{1} Yue Wu,\authormark{1} Yongxiang Chen,\authormark{1} Chongyu Lu,\authormark{1} and Yunjiang Rao\authormark{1}}

\address{\authormark{1}Key Laboratory of Optical Fiber Sensing
 and Communications, University of Electronic Science and Technology
  of China, Chengdu 611731, China\\
\authormark{2}Center for Information Geoscience, University of
 Electronic Science and Technology of China, Chengdu 611731, China
}
\email{\authormark{*}znwang@uestc.edu.cn} 



\begin{abstract}
Highly coherent wave is favorable for applications in which phase retrieval is necessary, yet a high coherent wave is prone to encounter Rayleigh fading phenomenon as it passes through a medium of random scatters. As an exemplary case, phase-sensitive optical time-domain reflectometry ($\Phi$-OTDR) utilizes coherent interference of  backscattering light along a fiber to achieve ultra-sensitive acoustic sensing, but sensing locations with fading won't be functional. Apart from the sensing domain, fading is also ubiquitous in optical imaging and wireless telecommunication, therefore it is of great interest. In this paper, we theoretically describe and experimentally verify how the fading phenomena in one-dimension optical scatters will be suppressed with arbitrary number of independent probing channels. We initially theoretically explained why fading would cause severe noise in the demodulated phase of $\Phi$-OTDR; then M-degree summation of incoherent scattered light-waves is studied for the purpose of eliminating fading. Finally, the gain of the retrieved phase signal-to-noise-ratio and its fluctuations were analytically derived and experimentally verified. This work provides a guideline for fading elimination in one-dimension optical scatters, and it also provides insight for optical imaging and wireless telecommunication.
\end{abstract}

\section{Introduction}
Fading is a ubiquitous phenomenon in the field of wireless telecommunication (three dimensions), optical imaging (two dimensions) and distributed optical fiber sensing (one dimension, 1-D). Fading phenomenon is characterized by a random attenuation of the signal. The signals may be reflected by various surfaces and reach the receiver via different paths. The received information is the sum of all the signals from varieties of paths. When they interfere with each other and out of phase, the fading phenomenon appeared. Such a phenomenon may lead to temporary failure of communication, deterioration of image quality and limitation of the sensor's credibility.

In order to comprehensively study fading phenomenon, reducing the dimension of the problem might be a wise way. The purpose of the optical imaging field is to obtain a speckle-free (non-fading-points) map with more emphasis on the intensity fluctuations \cite{CaoCUSTOM}. However, after the promotion of Phase Shift Keying (PSK) schemes, the effect of fading points on the demodulated phase is also a significant problem. Phase-sensitive optical time-domain reflectometry ($\Phi$-OTDR), which uses the randomly distributed Rayleigh scattering inside a fiber as the sensing mechanism, may be an ideal carrier \cite{newsensing}.

$\Phi$-OTDR, using an optical fiber (1-D scatters) to obtain acoustic information, needs both intensity information and phase information. Due to the maturity of the high-quality fiber manufacture process, there is no strong reflection point inside the fiber and the scattering elements are frozen. Besides, the external environment is more controllable compared with wireless telecommunication. Each resolvable segment of the fiber contains large number of randomly distributed Rayleigh scatters, which sufficiently satisfies the central limit theorem. Thus, the $\Phi$-OTDR system is suitable for theoretically analyzing high-dimensions fading phenomenon, such as the mode of tropospheric and ionospheric signal propagation as well as the effect of heavily built-up urban environments on radio signals \cite{FadingCom}.

The fading phenomenon in $\Phi$-OTDR comes from the randomly spatial non-uniform distribution of the refractive index \cite{hartog2017}. Fading points (extremely low intensity backscattering points) are detrimental for obtaining acoustic information in $\Phi$-OTDR system. Ref. \cite{eyal2016} has explained that the signal-noise ratio (SNR) of acoustic detection depends on intensities of the sampled points. At the fading points, the intensity noise will bring intolerably large noise after phase demodulation. That paper also gives an analytical formula for the mean fluctuation of retrieved phase $SNR_{\phi}$. The fluctuation of  $SNR_{\phi}$ was characterized by the coefficient of variation ($C_{V}$), which is as high as 0.8944 for traditional single frequency ${\Phi}$-OTDR \cite{eyal2017}. A similar definition exists in the field of optical imaging, namely speckle contrast, but it is used to describe intensity fluctuation.

In order to reduce fading points, a straightforward thought is to add M statistically independent field components (M degrees of freedom), such as orthogonal frequency division multiplexing (OFDM) \cite{goldsmith2005wireless} used in wireless telecommunication, multi-mode incoherent light \cite{cao2012} used in optical imaging and inner-pulse frequency-division method \cite{chen2017} used in DOFS. For $\Phi$-OTDR, Ref. \cite{zhou2013} discussed the rationality of using multi-frequency to eliminate fading phenomenon. The condition for two probe signals to be statistically independent is that their frequencies differ by at least the inverse pulse duration \cite{mermelstein2001rayleigh}. Since the backscattering signals is complex, it is not advisable to directly aggregate different frequency backscattering signals on the intensity, which causes phase mix-up. Ref. \cite{chen2017} proposed rotated-vector-sum method to aggregate backscattering signals with four different frequencies, which is proved to be effective. Besides, the relationship between the number of aggregated frequencies and $SNR{_{\phi }}$ of acoustic signal is shown in \cite{hartog2018use}. Based on the simulation results, the $SNR{_{\phi }}$ enhancement factor is closed to $\sqrt{M}$, which is similar to M averages on a single frequency intensity traces \cite{wang2018distributed}.  However,  to the best of authors' knowledge, no one has theoretically studied the statistical regularity of $SNR{_{\phi }}$ fluctuations after aggregating arbitrary degrees of freedom.

In this work, the fading problem in one dimension optical scatters has been studied in depth, using $\Phi$-OTDR as the platform. Firstly, the effect of intensity noise on the demodulated phase is derived, which is a completely different approach compared with previous literatures and more rigorous. Moreover, we introduced an expression for the amplitude distribution after aggregating M degrees of freedom. Then the $SNR_{\phi }$ gain and the level of $SNR_{\phi }$ fluctuations were also derived, and the results were verified by the simulations and experiments.

\section{The effect of intensity noise on the demodulated phase}

For the typical $\Phi$-OTDR system (single-carrier frequency, single-polarization pulses probing a single-mode fiber), the in-phase and quadrature outputs, i.e., ${i}$ and ${q}$, can be represented by (1) \cite{wang2016coherent}, which are asymptotically Gaussian distribution with zero mean and variance ${{\sigma }^{2}}$ without considering noise \cite{healey1987statistics}.

\begin{equation}
  \left\{ 
    \begin{aligned}
     i\left( k,t \right)\ =\ {{A}_{k,t}}\cos &\left[ \left( {{\omega }_{s}}-{{\omega }_{LO}} \right)t+{{\phi }_{k,t}} \right]+{{n}_{i}}\left( k,t \right) \\ 
    q\left( k,t \right)\ =\ {{A}_{k,t}}\sin &\left[ \left( {{\omega }_{s}}-{{\omega }_{LO}} \right)t+{{\phi }_{k,t}} \right]+{{n}_{q}}\left( k,t \right)
  \end{aligned}
  \right.
\end{equation}

In (1), $t$ stands for the fast time axis, indirectly representing the position of the fiber; $k$ stands for the slow time axis, indicating the changes of the information obtained at the same point after each pulse; ${{\omega }_{s}}$ is the angular frequency of signals while ${{\omega }_{LO}}$ is the angular frequency of local oscillator light; ${{\phi }_{k,t}}$ is the phase changes caused by external acoustic signals; ${{n}_{i}}$ and ${{n}_{q}}$ represent uncorrelated Gaussian white noise with zero mean and the same variance $\sigma _{n}^{2}$. It should be noted that for conveniently exploring the phase noise caused by intensity noise on the phase demodulation process, only intensity noise is considered. $A$ is the amplitude of backscattering signals. In a relatively short period of time, the amplitude (${{\left. {{A}_{k,t}} \right|}_{t=t'}}$) can be deemed as invariant \cite{hartog2017}.

Therefore, at the slow time axis, ${n_{i}}$ and ${n_{q}}$ are two linear independent quantities, and the amplitude of backscattering signal is invariant. Two parameters ${x_{k,t'}}$ and ${y_{k,t'}}$ whose definition are ${{n_{i}}/{A\left( t' \right)}}$ and ${n_q}/{A\left( t' \right)}$, respectively, are introduced for simplification. ${x}$ and $y$ follow the two-dimensional Gaussian distribution with variance ${{\delta }^{2}}\left( t' \right) = {\sigma _n^{2}}/{{A^{2}}\left( t' \right)}$, and the joint probability density of them is

\begin{equation}
  {{P}_{x,y}}(x,y)=\frac{1}{2\pi {{\delta }^{2}}}\exp [-\frac{1}{2}(\frac{{{x}^{2}}}{{{\delta }^{2}}}+\frac{{{y}^{2}}}{{{\delta }^{2}}})]
\end{equation}

The intensity noise existed in $i/q$ outputs can have impact on the demodulation phase signals, that is to say, the intensity noise transfers to phase noise after phase demodulation. The phase noise can be expressed as:

\begin{equation}
  \Delta \left( k,t' \right)=\arctan \left( \frac{\sin \Phi \left( k,t' \right)+\text{x}}{\cos \Phi \left( k,t' \right)+\text{y}} \right)-\Phi \left( k,t' \right)
\end{equation}
where $\Phi (k,t')=\left( {{\omega }_{S}}-{{\omega }_{LO}} \right)t'+{{\phi }_{k,t'}}$ is the sum of a constant and the external acoustic signal. Using Jacobi determinant, Eq. (2) and (3) can be integrated into the probability density function (PDF) of phase noise, as shown in Eq. (4). When the normalized intensity noise ${{\delta }^{2}}$ is small, the distribution of $\Delta \left( k,t' \right)$ is concentrated near 0, i.e., $sin\Delta \sim \Delta $ and $\cos \Delta \sim 1$.

\begin{equation}
  \begin{aligned}
      {{P}_{\Delta }}(\Delta)&=\int_{-\infty }^{+\infty }{{{P}_{x,y}}( x( y,\Delta  ),y )| \frac{\partial x}{\partial \Delta } |}dy\\ 
     &\approx \ \frac{1}{\delta \sqrt{2\pi }}\exp (-\frac{si{{n}^{2}}\Delta }{2{{\delta }^{2}}})\cos (\Delta )\approx \frac{1}{\sqrt{2\pi }\delta }\exp (-\frac{{{\Delta }^{2}}}{2{{\delta }^{2}}})
    \end{aligned}
\end{equation}

It should be noted that, the phase noise obeys Gaussian distribution approximatively, so the variance of the phase noise transferred from the intensity noise is about ${{\delta }^{2}}$. In other words, in the case of low intensity noise, i.e., ${{A}^{2}}\gg {{n}^{2}}$, the phase noise is inversely proportional to the  intensity of backscattering signal. For low intensity signal comparing with noise, the linear relationship is not valid. However, it is certain that the extremely low  intensity point will bring large phase noise and deteriorate the demodulated acoustic signal, which will cause the phenomenon of fading.

\section{The statistical regularity of retrieved phase $SNR_{\phi }$}

\subsection{Theory}

Due to the phase accumulation with the fiber length, it is necessary to use a differential process to obtain external acoustic signals. Therefore the $SNR_{\phi }$ of acoustic signal is associated with backscatter power of two points for acquiring differential phase, which is given by \cite{eyal2016}

\begin{equation}
  SNR_{\phi }=\frac{\sigma _{\phi }^{2}}{\sigma _{n}^{2}\left[ {1}/{{{A}^{2}}({{\text{t}}_{1}})+{1}/{{{A}^{2}}({{\text{t}}_{2}})}} \right]}
\end{equation}
where $\sigma _{\phi }^{2}$ is the variance of acoustic signal. In order to intuitively express the process of fading point reduction, calculating the mean and variance of $SNR_{\phi }$ is important, which depends on the amplitude distribution of two differential points.

Unlike intensity superposition \cite{Shimizu1992Characteristics}, the aggregation of complex signals is related to the signal angle. When the signal angles are all the same, the stacking gain is maximum, which is the same as the amplitude summation. The amplitude distribution for single pulse backscattering signals is Rayleigh distributed. A relatively simple and widely used small argument approximation for obtaining Rayleigh sum PDF was adopted in \cite{Beaulieu1990}. Although this approximation is not accurate enough, the error has little effect on subsequent calculations and the expression is concise. Approximate equation of the Rayleigh sum PDF is:

\begin{equation}
  \begin{aligned}
    {f_{sR}}(x)\ &=\ \frac{{{x}^{2M-1}}{{e}^{-\frac{{{x}^{2}}}{Mb}}}}{{{2}^{M-1}}{b^{M}}( M-1 )!{{M}^{M}}} \\
    b\ &=\ \frac{{{\sigma }^{2}}}{M}{{[ ( 2M-1 )!! ]}^{1/M}}
  \end{aligned}
\end{equation}
where $\left( 2M-1 \right)!!=(2M-1)(2M-3)\cdots 3\cdot 1$. It is assumed that each signal has the same intensity, namely $\left\langle A_{1}^{2} \right\rangle =\left\langle A_{2}^{2} \right\rangle =\cdots =\left\langle A_{M}^{2} \right\rangle =2{{\sigma }^{2}}$. Taking into account the intensity noise on each frequency trace, the intensity noise becomes $\sigma _{^{n,M}}^{2}=M\sigma _{^{n}}^{2}$ after overlapping M independent backscattering signals. Thus, the mean $SNR_{\phi }$ can be numerically calculated as:

\begin{equation}
  \begin{aligned}
    \mu \left( SNR_{\phi } \right)&=\iint{\left[ {{f}_{sR}}\left( A\left( {{t}_{1}} \right) \right)\frac{\sigma _{\phi }^{2}}{\sigma _{n,M}^{2}\left[ {1}/{{{A}^{2}}({{t}_{1}})+{1}/{{{A}^{2}}({{t}_{2}})}\;}\; \right]}{{f}_{sR}}\left( A\left( {{t}_{2}} \right) \right) \right]}dA\left( {{t}_{1}} \right)dA\left( {{t}_{2}} \right) \\ 
   & =\frac{2M{{\left( \left( -1+2M \right)!! \right)}^{1/M}}}{\left( 1+2M \right)}\frac{{{\sigma }^{2}}\sigma _{\phi }^{2}}{\sigma _{n}^{2}}\ \ \ \ \ \ \ \ \ \ \ \ =\frac{2}{3}{{K}^{2}}\sigma _{\phi }^{2}SNR_{intensity}
  \end{aligned}
\end{equation}
where $SN{{R}_{intensity}}\equiv {{{\sigma }^{2}}}/{\sigma _{n}^{2}}$ is intensity SNR for single frequency backscattering signal and the $K$ is the gain of $SNR_{\phi }$. However, the expression of gain $K$ is very complicated with double factorial. In order to simplify the expression, Stirling’s approximation \cite{Dutka1991} is adopted, which is a good formula for accurately estimating factorials. The gain $K$ can be approximated as:

\begin{equation}
  {{K}^{2}}\ =\ \frac{3M{{\left( \left( -1+2M \right)!! \right)}^{1/M}}}{\left( 1+2M \right)}\ \approx\  \frac{6{{M}^{2}}}{\left( 2M+1 \right)e}
\end{equation}
where $e$ is the mathematical constant. Notably, for larger value of $M$, gain $K$ approaches to $\sqrt{3M/e}\approx 1.05\sqrt{M}$.

In order to reflect the effect of reducing fading, the level of $SNR_{\phi }$ fluctuations can also be an important indicator. However, the statistical distribution of $SNR_{\phi }$ is particularly wide, and it is unsuitable to use the variance to measure the $SNR_{\phi }$ fluctuation. Using the $C_{V}$ to measure it may be more lucid, which can be expressed as

\begin{equation}
  {{C}_{V}}\left( SN{{R}_{\phi }} \right)=\frac{\sigma \left( SN{{R}_{\phi }} \right)}{\mu \left( SN{{R}_{\phi }} \right)}=\sqrt{\frac{2{{M}^{2}}+5M+1}{4{{M}^{3}}+6{{M}^{2}}}}
\end{equation}
where $\sigma \left( SNR_{\phi } \right)$ is the standard deviation of $SNR_{\phi }$. $\sigma (SNR_{\phi })\equiv \sqrt{\mu \left( SNR_{\phi }^{2} \right)-{{\mu }^{2}}\left( SN{{R}_{\phi }} \right)}$, where $\mu \left( SNR_{\phi }^{2} \right)$ can be obtained with the same method as Eq. (7).

\subsection{Experiments and simulations}

\begin{figure}[htb!]
  \centering\includegraphics[width=11cm]{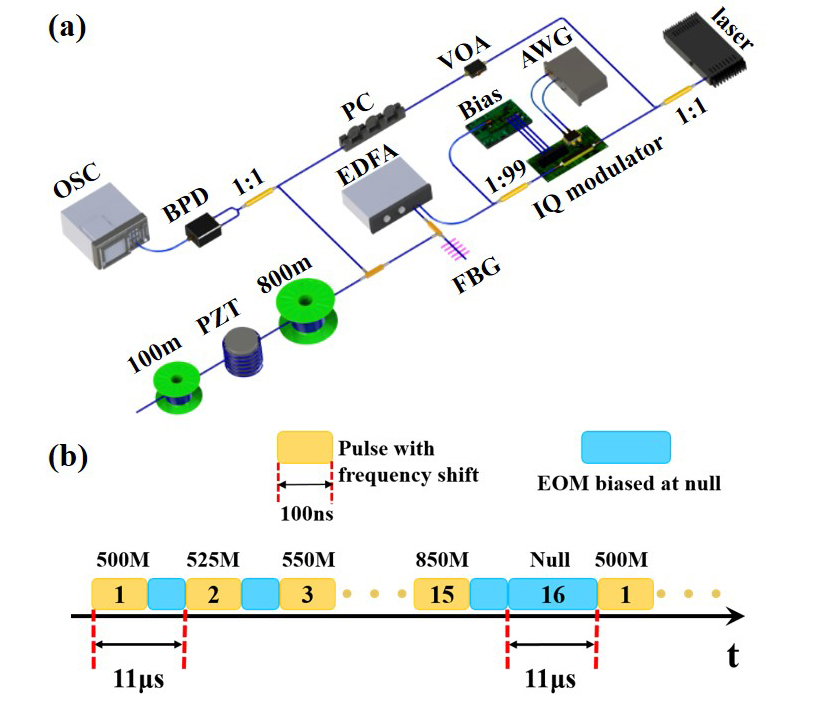}
  \caption{Experimental setup of $\Phi$-OTDR. (a) AWG, arbitrary waveform generator; VOA, variable optical attenuator; EDFA, Erbium-doped fiber amplifier; FBG, fiber Bragg grating; PC, polarization controller. (b) 15 groups of differential frequency-shift pulse generated by IQ modulator and the 16th group is to measure the intensity noise. }
\end{figure}

The experiment setup shown in Fig. 1 is to verify the impact of aggregating M degrees of freedom on eliminating fading. An ultra-narrow linewidth laser with small phase noise is used, and the output is split into two branches by a 1/1 coupler, namely the signal branch and the local oscillator (LO) branch. The signal branch is modulated by an IQ modulator, generated frequency shift pulse with 100ns width. After amplified by EDFA, the pulse is injected into the sensing fiber through a circulator. A piezoelectric ceramic transducer (PZT) wrapped with 12.7 m fiber is inserted between the 800 m and 100m fiber sections, applied with 100 Hz sinusoidal perturbation. The Rayleigh scattering signal is injected into 2$\times$2 coupler together with the LO. The outputs of 2$\times$2 coupler are converted into electrical intensity signals by a 1.6 GHz balanced photo-detector (BPD), and then sampled by an oscilloscope (OSC) with 2 GS/s sampling rate. The initial frequency shift of the pulse is 500 MHz. After a round-trip time of the pulse in the fiber, another pulse with a larger frequency shift is applied, as shown in Fig. 1(b). The frequency shift interval is 25 MHz to ensure that each group of trace statistically independent. There are 15 groups of pulses with different frequency shifting from 500 MHz to 850 MHz, and another round trip time is reserved for measure the intensity noise, during which the IQ modulator stays at the null point. This method helps detect the leakage of light from the IQ modulator and the noise contributed by the EDFA.

\begin{figure}[htb!]
  \centering\includegraphics[width=13cm]{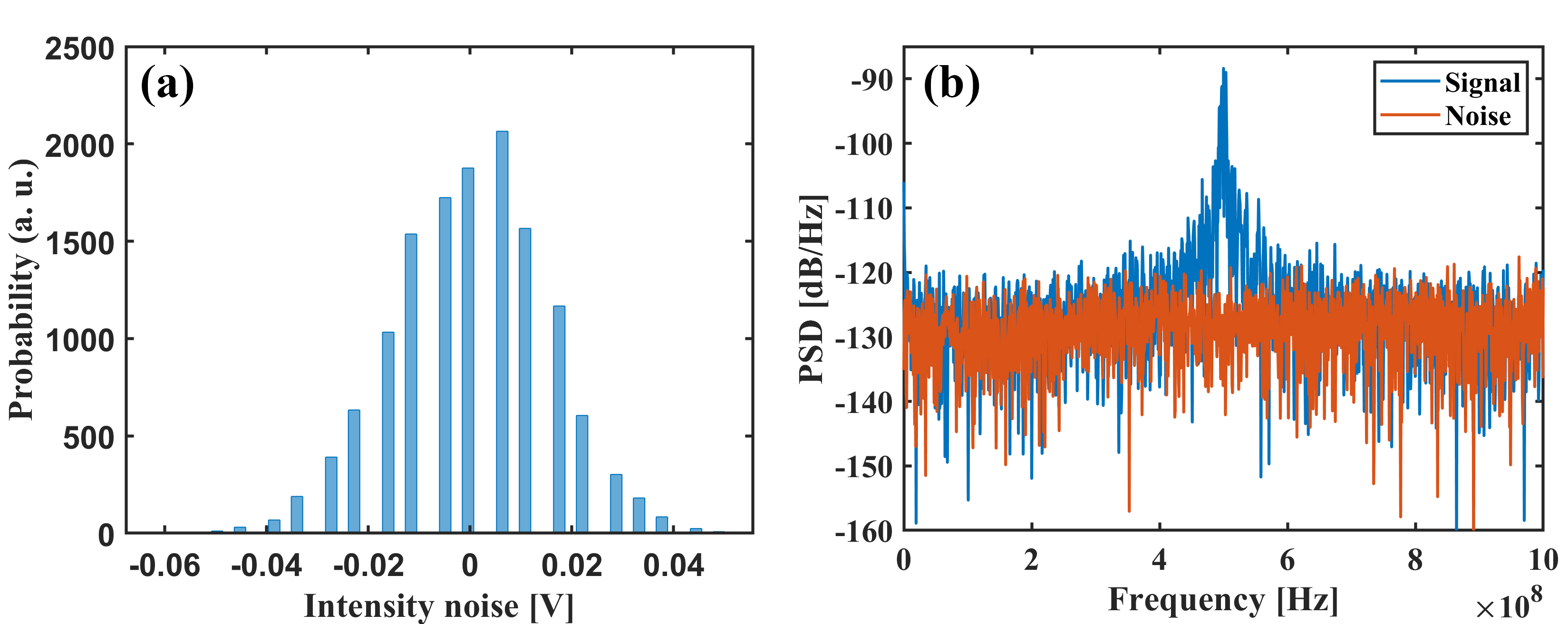}
  \caption{Statistical results of intensity noise. (a) the probability distribution. (b) PSD (blue line: signal; read line: noise).}
\end{figure}

From the experimental results, the main noise of the system comes from the detector and the local oscillator. The PDF and power spectral density (PSD) of the real part of the signal noise are shown in Fig. 2(a) and (b), respectively. The probability distribution of noise is approximately Gaussian, and the gap in it is caused by insufficient vertical resolution of the oscilloscope. In addition, the PSD of the noise is flat, so it can be considered as Gaussian white noise. The PSD of measured noise coincides with the background noise of the signal PSD, which indicates that measuring intensity noise in this way is reasonable.

\begin{figure}[hb!]
  \centering\includegraphics[width=13.3cm]{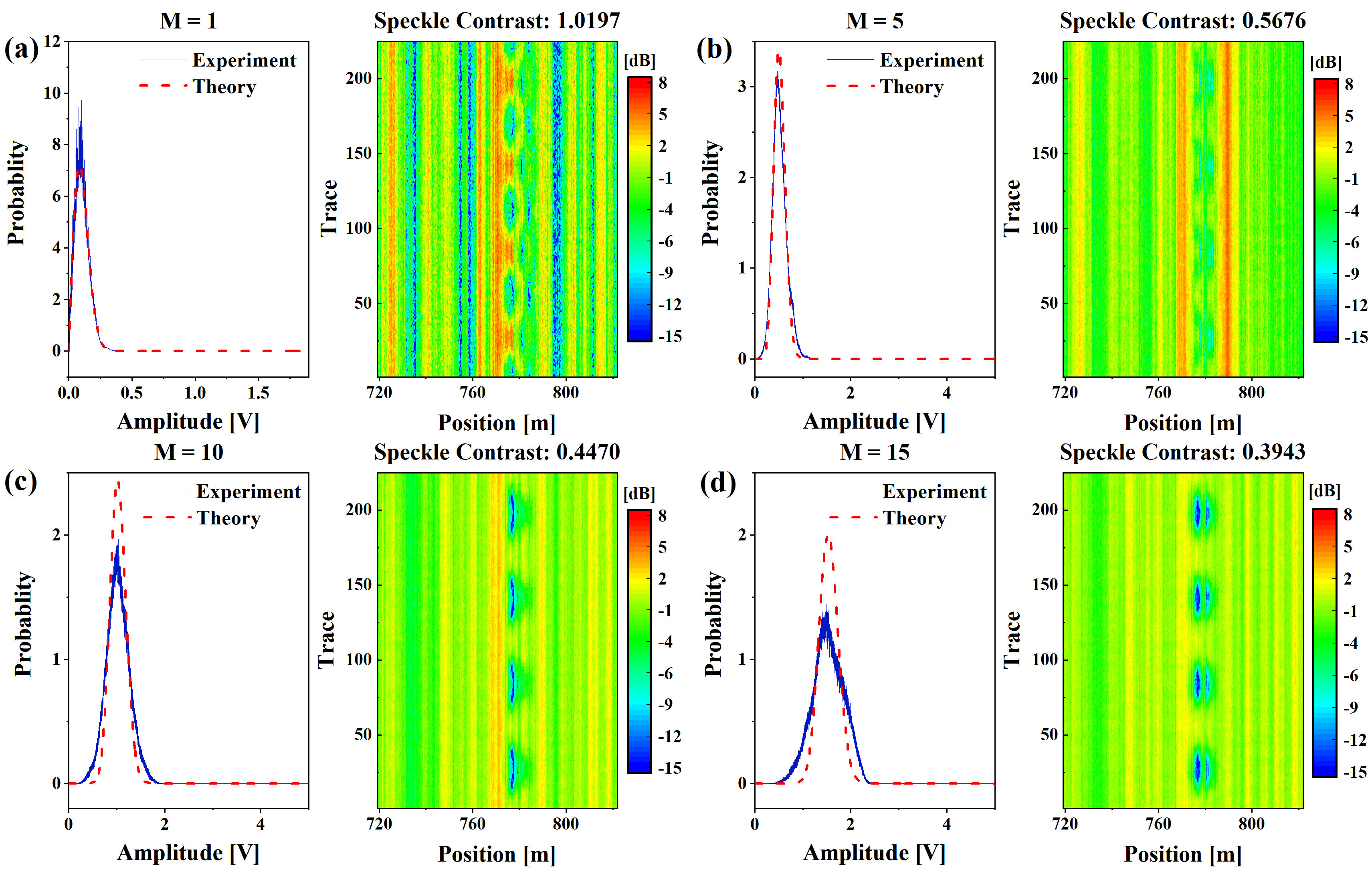}
  \caption{The amplitude distribution (left side) and normalized time-dependent backsactter profile (right side) after aggregating M uncorrelated backscattering signals.}
\end{figure}

In order to effectively aggregate scattered signals with different carrier frequencies and verify the aforementioned theoretical results, phase alignment is required. We used rotated-vector-sum method to aggregate each backscattering signals and the normalized results of time-dependent optical backscattering signal was shown in Fig. 3 (right-hand side of each sub-figure). As the number of superimposed frequencies increases, the intensity fluctuations gradually decrease, which can be characterized by the speckle contrast. The value of it from each ${{\sigma }_{I}}$ image was calculated as ${{{\sigma }_{I}}}/{\left\langle I \right\rangle }$, where  is the standard deviation of the intensity and $\left\langle I \right\rangle $ is the mean intensity. After superimposing 15 degrees of freedom in $\Phi$-OTDR, the speckle contrast decreased from ~1.02 to ~0.39 and the disturbed region is more and more distinguished. Besides, the amplitude distribution after aggregating is plotted in Fig. 3 (left-hand side of each sub-figure). The theoretical curve was based on Eq. (6). The value of $ {{\sigma }^{2}}$ in it was taken from the experiment (${{\sigma }^{2}}$ = 0.014076). In Fig.3, experimental and theoretical results will have a few differences after aggregating more than 5 degrees of freedom. This can be attributed to the fact that the gain spectrum of detector is not flat, then the strength of each scattered signal will be slightly different, as a result the PDF of the amplitude gradually becomes broadened. Therefore, it can be expected that the value of experimental $C_{V}$ will be slightly higher than the theoretical value. 

\begin{figure}[htb!]
  \centering\includegraphics[width=10cm]{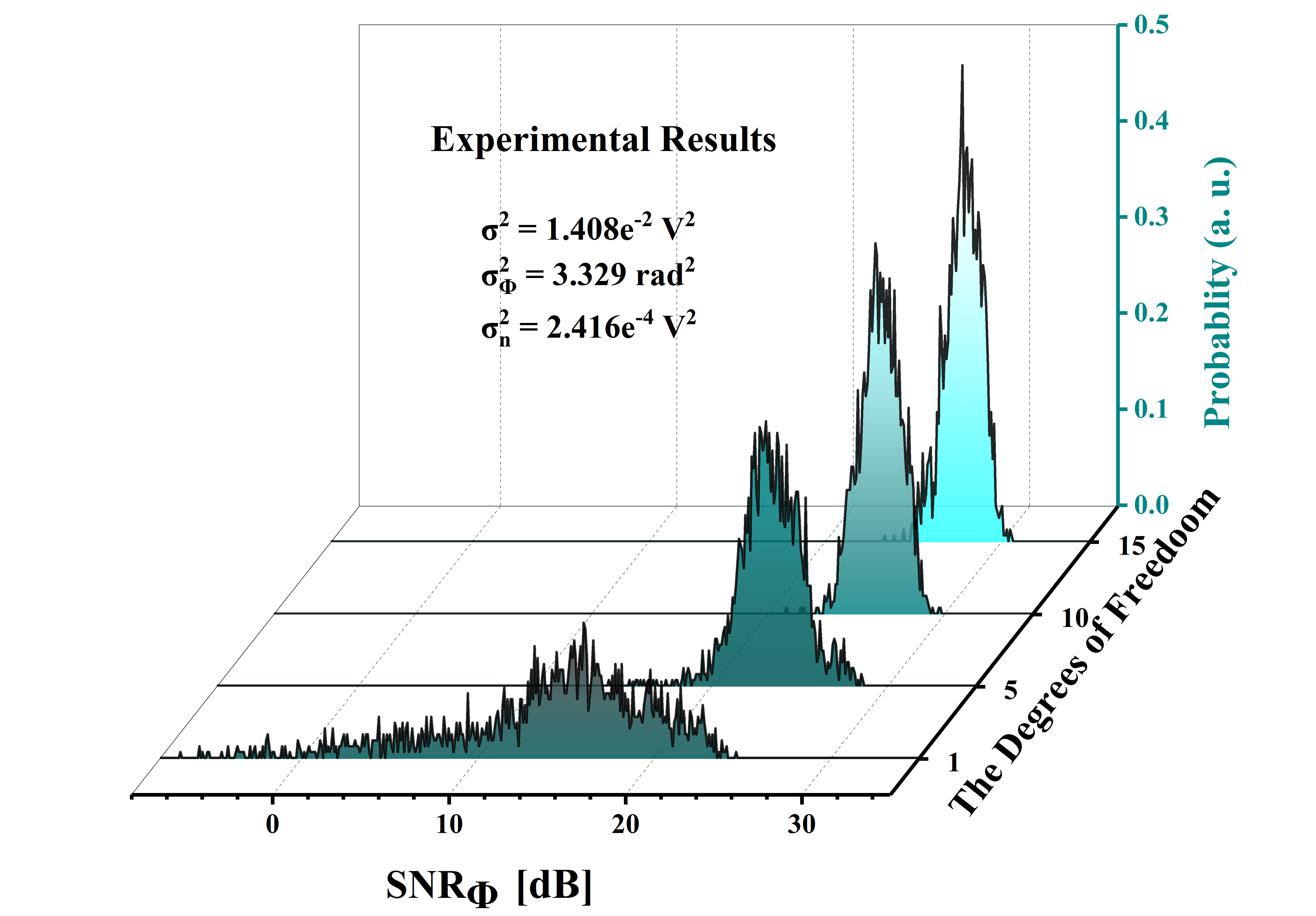}
  \caption{M degrees of freedom v.s. the PDF of $SNR_{\phi }$ from experimental results.}
\end{figure}

To obtain the mean and the $C_{V}$ of $SNR_{\phi }$, the measurements were repeated over 20 times. The gauge length is set to be 40 m, which is much larger than the PZT disturbance length (12.7 m). This allows us to get more than a hundred sets of PZT disturbance signals after one measurement, which is conducive to statistic $SNR_{\phi }$. The distribution of $SNR_{\phi }$ is plotted in Fig. 4. The parameters in Eq. (7), obtained from experiments, are also listed inside the figure. As the degrees of freedom increases, the distribution of $SNR_{\phi }$ became sharper and average $SNR_{\phi }$ increased.

\begin{figure}[htb!]
  \centering\includegraphics[width=13.3cm]{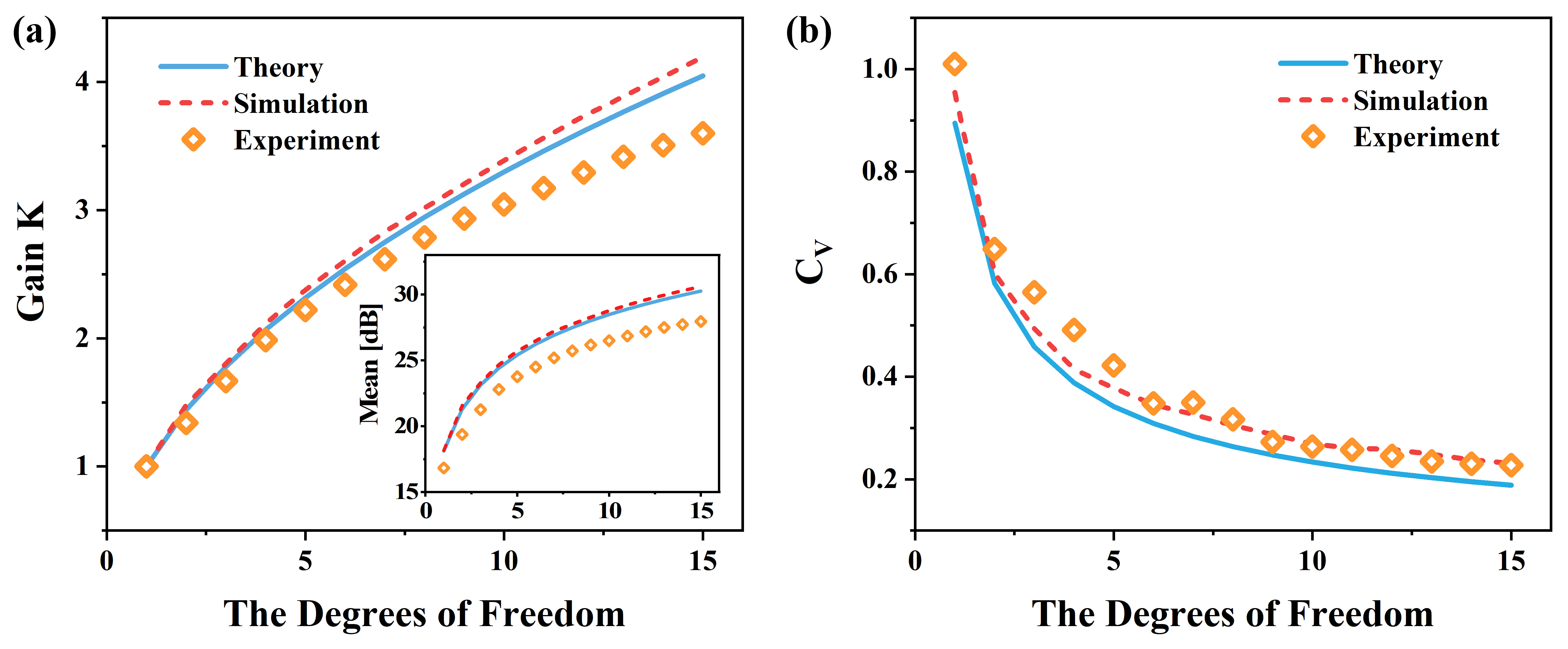}
  \caption{(a) Gain $K$, mean (inset in (a)) and (b) coefficient of variation of $SNR_{\phi }$ v.s. M degrees of freedom.}
\end{figure}

A set of simulations is also used to verify the correctness of the theory. The fiber parameters are based on Ref. \cite{Masoudi2017model}. The scatters in the fiber are randomly distributed. The perturbation of the fiber is introduced by changing the separation among the affected scatters. Besides, the intensity noise is fixed in each simulation and is consistent with the $SNR_{intensity}$ in the experiment. For statistic study, the positions of the scattering elements are considered to be fixed in the same simulation, but they are completely different among the simulations. The illustration in Fig. 5(a) shows the results of mean value of $SNR_{\phi }$ obtained by theory (blue), simulations (red) and experiments (yellow). Mean value of $SNR_{\phi }$ from experiments are lower than the theory and simulation, which may come from the polarization effects and Hilbert transform error in this system. However, the experimental results of gain K at the low degrees of freedom ($M\le 5$) is closed to theoretical and emulational results (plotted in Fig. 5(a)). The subsequent deviation of the gain $K$ (when $M>5$) is related to non-flat gain spectrum of the detector. In addition, the changes of $C_{V}$ value have been drawn in Fig. 5(b). Since the scattering elements are completely different in each simulation but the intensity noise is fixed, it results in different $SNR_{intensity}$ in each simulation, which is more obvious in experiments. But the fluctuation range of the $SNR_{intensity}$ is only 0.5 dB in the experiments, which is much lower than the $SNR_{\phi }$ fluctuation. Due to slight fluctuation of $SNR_{intensity}$ results the experimental and emulational $C_{V}$ value are slightly larger than the theoretical value, but the tendency is basically the same.

\section{Conclusion}
We studied the fading phenomenon in one-dimension optical scatters with statistical analysis. The reason of occurring fading phenomenon at low intensity point was mathematically explained. Moreover, the quantitative relationship between the degree of freedom and fading phenomenon was analyzed in detail. This work provides a lucid guideline for the choice of the degree of freedom to eliminate fading in 1-D optical scatters, which is crucial of fiber sensing based on Rayleigh scattering, and it would be also beneficial for optical imaging and wireless telecommunication domains.

\section*{Funding}

This work is supported by Natural Science Foundation of China (41527805, 61731006), Sichuan Youth Science and Technology Foundation (2016JQ0034), Guofang Keji Chuangxin Tequ, and the 111 project (B14039).

\section*{Acknowledgment}
The authors thank Anderson S. L. Gomes in Universidade Federal de Pernambuco for helpful discussions.



\end{document}